\documentstyle[aps,prl,preprint,floats,epsfig]{revtex}

\begin{document}

\def\etal{{\em et al.}}
\newcommand{\mev   }{\mbox{\rm MeV}}
\newcommand{\mevc  }{\mbox{\rm MeV/$c$}}
\newcommand{\mevcsq}{\mbox{\rm MeV/$c^2$}}
\newcommand{\dedx}{\mbox{$dE/dx$}}
\preprint{\tighten\vbox{\hbox{CLNS 98/1581}
\hbox{CLEO 98-13}
}}
\title
{\bf {\Large Observation of Two Narrow States Decaying Into $\Xi_c^{+}\gamma$ and $\Xi_c^{0}\gamma$}}  

\author{CLEO Collaboration}
\date{\today}

\maketitle
\tighten

\begin{abstract}

 We report the first observation of two narrow charmed strange baryons decaying
to $\Xi_c^{+}\gamma $ and $\Xi_c^0\gamma $, respectively, using data from 
the CLEO II detector at CESR.  We interpret the observed signals as the 
$\Xi_c^{+\prime}(c\{su\})$ and $\Xi_c^{0\prime}(c\{sd\})$, the symmetric 
partners of the well-established antisymmetric $\Xi_c^{+}(c[su])$ and 
$\Xi _c^{0}(c[sd])$.  The mass differences $M(\Xi_c^{+\prime })-M(\Xi _c^{+})$
and $M(\Xi _c^{0\prime})-M(\Xi _c^0)$ are measured
to be $107.8\pm 1.7\pm 2.5$ \ and $107.0\pm 1.4\pm 2.5$ \mevcsq,
respectively.  
\end{abstract}
\normalsize
\newpage
{
\renewcommand{\thefootnote}{\fnsymbol{footnote}}
\begin{center}
C.~P.~Jessop,$^{1}$ K.~Lingel,$^{1}$ H.~Marsiske,$^{1}$
M.~L.~Perl,$^{1}$ V.~Savinov,$^{1}$ D.~Ugolini,$^{1}$
X.~Zhou,$^{1}$
T.~E.~Coan,$^{2}$ V.~Fadeyev,$^{2}$ I.~Korolkov,$^{2}$
% ...
\end{center}
 
\small
\begin{center}
$^{1}${Stanford Linear Accelerator Center, Stanford University, Stanford,
California 94309}\\
$^{2}${Southern Methodist University, Dallas, Texas 75275}\\
% ...
\end{center}
\setcounter{footnote}{0}
}
\newpage

CLEO\cite{alam,avery} and other experimental groups\cite
{biagi,coteus,barlag,albrecht} have previously reported the observation of
the $J^P=(\frac 12)^{+}$\ ground states $\Xi _c^0$\ $(c[sd])$ and $\Xi _c^{+}
$\ $(c[su])$ baryons, where [su] and [sd] denote the antisymmetric nature of
their wave functions with respect to interchange of the light quarks.
The partners of the above
charmed strange baryons are the $\Xi _c^{0\prime }$\ $(c\{sd\})$ and $\Xi
_c^{+\prime }$\ $(c\{su\})$, where $\{sd\}$ and $\{su\}$ specify that the wave
functions are symmetric with respect to interchange of the light quarks. In
this report we present the first observation of the $\Xi_c^{\prime}$ states\cite{notation}. 
 The $J^{P}=\frac{3}{2}^{+}$ spin-excited states $\Xi_c^{*0}$\ and
$\Xi_c^{*+}$, recently observed by CLEO \cite{avery95,gibbons96}, have spin-1 light
diquarks like the $\Xi_c^{\prime}$, in contrast to spin-0 light diquarks 
in the $\Xi_c$ states.
 The mass splitting $M(\Xi _c^{\prime })-M(\Xi _c)$
\cite{franklin,boyd,maltman,pirjol,ron,savage2,falk,jenkins}
is expected to be in the range of $100-114$ \mevcsq . 
With such a mass difference, the transition $\Xi_c^{\prime }\rightarrow \Xi _c\pi $ is kinematically 
forbidden, allowing only the decay $\Xi _c^{\prime }\ \rightarrow $ 
$\Xi _c\gamma $. 
The above theoretical models also predict the mass difference 
$M(\Xi _c^{*}) - M(\Xi_c^{\prime})$ to be about 60-70 \mevcsq.

The data used in this analysis was collected with the CLEO II detector \cite
{kubota} operating at CESR, and corresponds to an integrated luminosity of
4.96 fb$^{-1}$\ from the $\Upsilon \left( 4S\right) $\ resonance and
continuum region at energies just below it.  The charmed strange baryon $\Xi_c^0$ was
reconstructed in the decay modes $\Xi ^{-}\pi ^{+},\Xi^{-}\pi ^{+}\pi ^0,
\Xi ^0\pi ^{+}\pi ^{-}$ and $\Omega ^{-}K^{+}$, and $\Xi_c^{+}$ in the decay 
modes $\Xi ^{-}\pi ^{+}\pi ^{+}$ and $\Xi ^0\pi ^{+}\pi ^0$\cite{avery95,gibbons96,edwards}. 
In all cases, the signal area above the combinatorial
background is found by fitting to the sum of one or more Gaussian functions
with widths fixed at Monte Carlo predicted values, and a low-order Chebychev
polynomial. Where particle identification is used, a joint probability for the 
pion, kaon, or proton hypothesis is defined using measurements of specific 
ionization (\dedx) in the wire drift chambers and time-of-flight in the 
scintillation counters. A charged track is defined to be consistent with a 
particular mass hypothesis if the corresponding probability is greater than 
0.1\%.

Charmed baryons can be produced from either secondary decays of $B$ mesons
or directly from $e^{+}e^{-}\,$annihilations to $c\overline{c}$ jets. We
define $x_p$ and $x_p^{\prime }$ as the scaled momentum of the $\Xi _c\,$and 
$\Xi _c^{\prime }$, respectively. Here $x_p=p/p_{\max }$; $p$ is the
momentum of the charmed baryon, $p_{\max }=\sqrt{E_b^2-M^2}$, $E_b$ is the
beam energy and $M$ is the mass of the charmed baryon being considered.
Charmed baryons produced from $B$ decays are kinematically limited to $%
x_p<0.4$, while $(60-70)\%$ of those produced from the continuum have $x_p>0.4$.
To reduce random combinatorial background, we apply a mode-dependent cut of 
$x_p>0.5-0.6$, thus excluding charm baryons produced in $B$ decays.

We begin by reconstructing
\thinspace $\Lambda \rightarrow p\pi ^{-},$ $\Xi ^0\rightarrow \Lambda \pi ^0
$, $\Xi ^{-}\rightarrow \Lambda \pi ^{-}$, and $\Omega ^{-}\rightarrow \Lambda
K^{-}.$ 
We select hyperons by requiring the distance between the reconstructed secondary
decay vertex and the beam interaction point as measured in the plane 
perpendicular to the beam line, to be at least 2 mm for $\Lambda$ and  
$\Xi^{-}$, and 3 mm for $\Xi^{0}$, respectively.  No such cut is applied for $\Omega^{-}$.

Candidates for $\Lambda \rightarrow p\pi ^{-}$ decays\ are reconstructed
from pairs of oppositely charged tracks, assuming the higher momentum one to
be a proton and requiring it to be consistent with the proton hypothesis. 
The invariant mass of the combination is calculated using a three-dimensional 
vertex-constrained fit at the point of intersection. All $p\pi ^{-}$ 
combinations within 5 \mevcsq\ ($\approx$ 3 standard deviations $(\sigma)$) 
of the nominal mass are accepted as $\Lambda $\ candidates.

A $\Xi ^{-}$\ candidate vertex is reconstructed by
finding the intersection between a $\Lambda $\ candidate and $\pi ^{-}$
track, and requiring the $\Xi ^{-}$ direction to be consistent with coming
from the event vertex. A fit to the resultant distribution 
of $\Lambda \pi^{-}$\ invariant mass combinations 
yields a total of 11578 $\pm $ 125 reconstructed $\Xi ^{-}$\ candidates. All
such combinations within 5 \mevcsq\ $(\approx 3\sigma )$ of the nominal
mass are accepted as $\Xi ^{-}$ candidates. 

For $\Omega ^{-}$ reconstruction, we combine each $\Lambda $ candidate with
any negatively charged track that is consistent with the kaon hypothesis. 
The $\Omega ^{-}$ vertex is found using a procedure very
similar to that used for finding $\Xi ^{-}$. A fit to the distribution of 
$\Lambda K^{-}$\ invariant mass combinations yields 
a signal of 373 $\pm$ 32 events, and combinations within 5 \mevcsq\ of the 
nominal mass are selected as $\Omega ^{-}$\ candidates. 

The $\Xi ^0$\ candidates are reconstructed from $\Lambda $ and $\pi ^0$
pairs. Candidates for $\pi ^0$ are formed from pairs of photons detected in
the CsI calorimeter, with at least one photon coming from the barrel 
($| \cos \theta | < 0.7$) rather
than the endcap regions, where $\theta$ is the polar angle with respect 
to the $e^+$ direction. Only photons with energy greater than 50 MeV
and distinctly separated from charged tracks are used. As a first
approximation, the $\pi ^0$ mass is calculated assuming the event vertex to be
its point of origin. A $\Xi ^0$ vertex is then found from the
intersection of the $\Lambda $ and $\pi ^0$ directions. The mass and 
four-momentum of the $\pi ^0$ is recalculated assuming the $\Xi ^0$ vertex to 
be its origin. A new vertex is calculated using the new $\pi ^0$ and $\Lambda $
directions. A fit to the $\Lambda \pi^{0}$ mass distribution yields 
7568 $\pm$ 227 signal events, and all $\Lambda \pi ^0$ combinations within 
8 \mevcsq\ of the nominal mass are defined as $\Xi ^0$ candidates.

\begin{figure}[ht]
\unitlength 1.0in
\begin{picture}(6.0,6.0)(0.0,0.0)
\put(0.0,0.0){\psfig{file=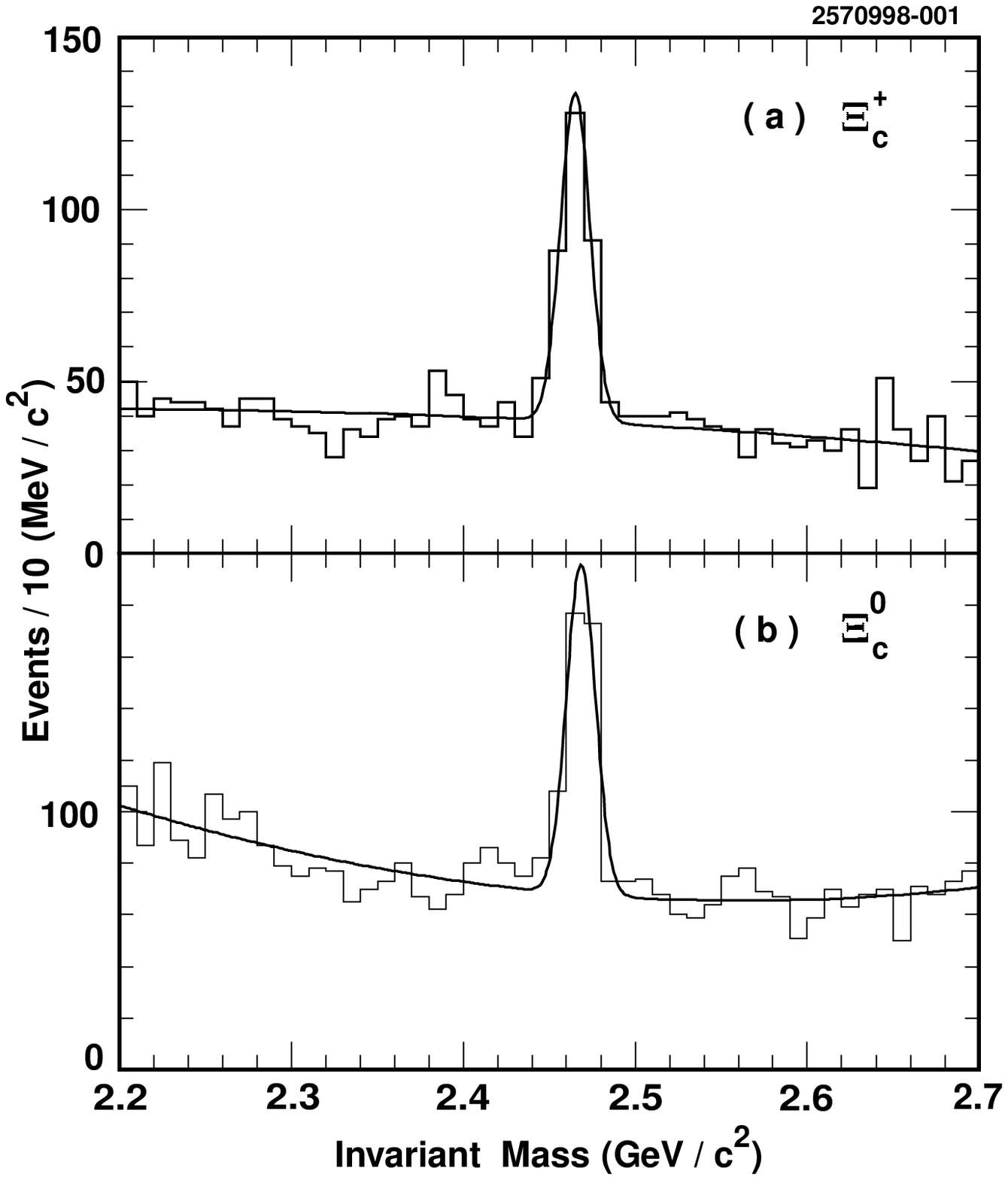,height=5.7in,bbllx=40,bblly=140,bburx=480,bbury=650}}
\end{picture}
\caption[]{(a) Summed invariant mass distributions for $\Xi^{-}\pi^{+}\pi^{+}$ 
and $\Xi^{0}\pi^{+}\pi^{0}$ combinations with $x_p>0.5$ and 0.6, respectively, and (b)  
for $\Xi^{-}\pi^{+}$, $\Xi^{-}\pi^{+}\pi^{0}$, $\Omega^{-}K^{+}$, and $\Xi^{0}\pi^{+}\pi^{-}$ combinations with
$x_p>$0.5, 0.5, 0.5 and 0.6, respectively.}
\label{fig:fgcscp}
\end{figure}

We first discuss the reconstruction of $\Xi _c^{+}$ candidates in the decay
modes $\Xi ^{-}\pi ^{+}\pi ^{+}\,$and $\Xi ^0\pi ^{+}\pi ^0.$ As presented
earlier, $\Xi ^{-}$ and $\Xi ^0$ candidates are combined with charged or
neutral pions which are consistent with originating from the event vertex.
In the case of the first decay mode, only charged tracks with momentum
greater than 100 \mevc\ are used. For the second decay mode, which has more
combinatorial background because of the $\pi ^0$, both the charged and
neutral pions are required to have momenta greater than 250 \mevc.
We form invariant mass distributions of $\Xi ^{-}\pi ^{+}\pi ^{+}$ 
combinations with $x_p>0.5$ and $\Xi ^0\pi ^{+}\pi ^0$ combinations 
with $x_p>0.6$.  Fitting these 
distributions with Monte Carlo predicted widths of 8.5 and 15 \mevcsq, respectively, 
we obtain yields of $(155\pm 15)$ and $(70\pm 14)$ signal events in these
two decay modes or a combined yield of $(225\pm 21)$. Combinations within $
2\sigma $ of the fitted peak masses in each decay mode are then selected as $
\Xi _c^{+}$ candidates.  The invariant mass distribution for the
summed combinations in both $\Xi_c^{+}$\ decay modes is shown in 
Fig. \ref{fig:fgcscp}(a). 

We reconstruct $\Xi _c^0$ in the four decay modes $\Xi ^{-}\pi ^{+}$, $\Xi
^{-}\pi ^{+}\pi ^0$, $\Omega ^{-}K^{+}$, and $\Xi ^0\pi ^{+}\pi ^{-}$. We
start with the hyperon candidates, which are defined according to procedures
discussed previously, and add charged tracks which are consistent with
coming from the event vertex. For the decay mode $\Xi ^{-}\pi ^{+}\pi ^0$,
we assume the photons used for reconstructing $\pi ^0\rightarrow \gamma
\gamma $ are coming from the event vertex. Only $\gamma \gamma $
combinations having invariant mass within 12.5 \mevcsq\ $(2.5\sigma )$ of
the nominal mass are used as $\pi ^0$ candidates. In the case of
$\Omega^{-}K^{+}$, we use only primary charged tracks consistent with the 
kaon hypothesis. Only combinations with $x_p>0.5$ are used in
the case of the first three decay modes; for the last decay mode, since the
combinatorial background is higher, a cut of $x_p>0.6$ is used. 
Fitting the invariant mass distributions corresponding to the decay modes
 $\Xi ^{-}\pi ^{+}$, $\Xi
^{-}\pi ^{+}\pi ^0$, $\Omega ^{-}K^{+}$, and $\Xi ^0\pi ^{+}\pi ^{-}$
with Monte Carlo predicted widths of $8,10,7$ and $12$
MeV/$c^2$,  we obtain yields of $(133\pm 41),(86\pm 13),(24\pm
5)$ and $(46\pm 10)$ signal events, respectively.  
This gives a combined $\Xi_c^{0}$ yield of ($289\pm44$) events.
The sum  of the four $\Xi_c^{0}$\ invariant mass distributions is shown in 
Fig. \ref{fig:fgcscp} (b).  

To search for $\Xi _c^{+\prime }$ and $\Xi _c^{0\prime }$, we start with
the $\Xi _c^{+}$ and $\Xi _c^0$ candidates reconstructed according to the
procedure described in the earlier sections. We then form $\Xi _c^{+}\gamma $
and $\Xi _c^0\gamma $ combinations using photons with energy greater than 
100 MeV. Only showers detected in the barrel CsI crystal calorimeter ($|\cos\theta| < 0.7$), 
with clear isolation from nearby charged tracks and shower fragments are
used as photon candidates. The lateral shower profile of the candidate
is required to be consistent with that of a photon. A photon is also rejected if
it is part of a good $\pi ^0$ candidate, as defined in the section on $\Xi
_c^0$ reconstruction. About $(30-50)\%$ of photons from
$\Xi_c^{\prime}$\ 
are lost due to this veto. Instead of plotting the $\Xi _c\gamma $ 
invariant mass combinations, we plot the mass difference 
$\Delta M=M(\Xi _c\gamma )-M(\Xi _c)$, 
which has better mass resolution as the errors from $\Xi_c$
reconstruction are common to both terms and therefore cancel. In
plotting the $\Delta M\,$
distributions, the $x_p$ cut on $\Xi _c$ reconstruction is removed and
instead we place a cut on $x_p^{\prime }$, the $x_p$ of the $\Xi_c\gamma$ combination.  Final states including $\Xi_0$ have larger combinatorial
backgrounds.  We therefore require $x_p^{\prime} > 0.6$ for these states
and $x_p^{\prime} > 0.5$ for all other final states.

\begin{figure}[ht]
\unitlength 1.0in
\begin{picture}(6.0,6.0)(0.0,0.0)
\put(0.0,0.0){\psfig{file=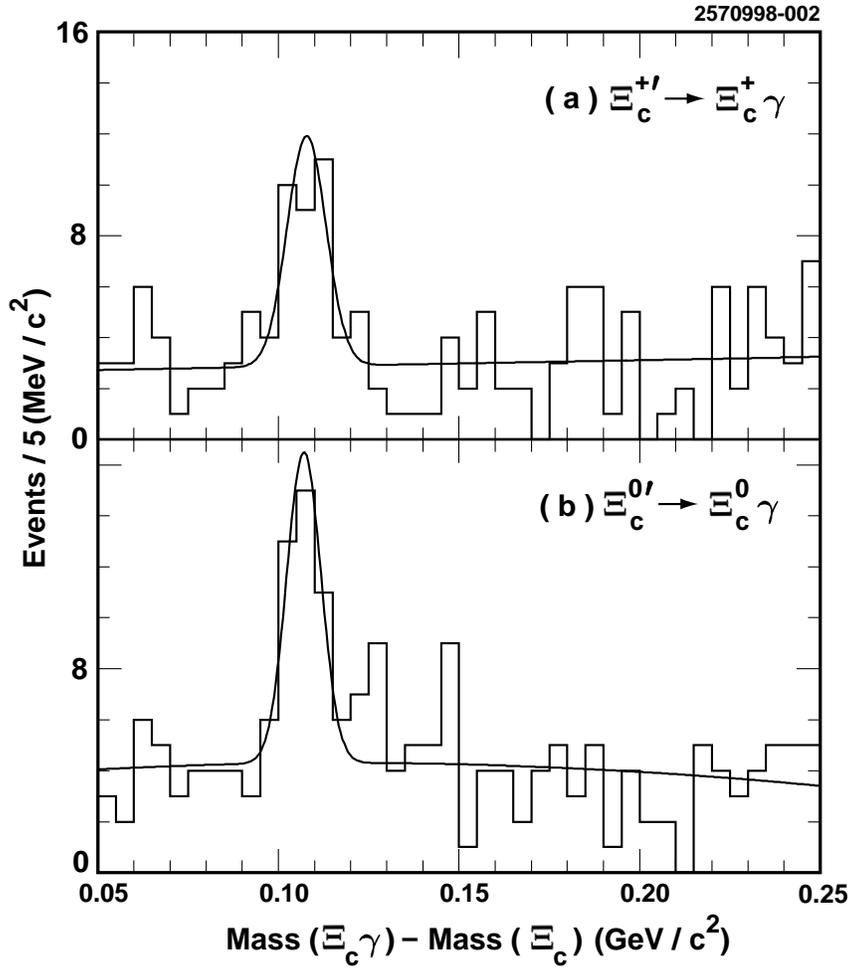,height=5.7in,bbllx=40,bblly=140,bburx=480,bbury=650}}
\end{picture}
\caption[]{ Invariant mass difference $\Delta M(\Xi_c\gamma - \Xi_c)$
distributions for $\Xi_c^+\gamma$ and $\Xi_c^0\gamma$, where contributions from
the different $\Xi_c$ decay modes have been summed in each case.}
\label{fig:fgdmcomb}
\end{figure}

Fitting the mass difference $\Delta
M^{+}=M(\Xi _c^{+}\gamma )-M(\Xi _c^{+})$ distributions corresponding to the
two $\Xi _c^{+}$ decay modes used in the analysis, we obtain $
(16.1\pm 5.1)$ and $(7.5\pm 3.6)$ signal events, respectively. Similarly, fits 
to the mass difference $\Delta M^0=M(\Xi _c^0\gamma )-M(\Xi _c^0)$ distributions corresponding
to the four $\Xi _c^0$ decay modes separately yield signal areas of $%
(7.0\pm 4.0)$, $(11.6\pm 4.4)$, $(3.8\pm 2.0)$, and $(6.0\pm 3.3\,)$ events, 
respectively. It may be noted that there is at least one mode in each case with
an enhancement of $3\sigma $ statistical significance and corroborating
enhancements in the other decay modes in the mass difference region around 108 \mevcsq. 
Fig.~\ref{fig:fgdmcomb} (a) and (b) show the combined mass difference 
distributions for the  $\Xi_c^{+}\gamma$ and $\Xi_c^{0}\gamma$ 
combinations, respectively, where the contributions from the different decay 
modes have been summed. The distributions are fitted with 
widths fixed at the Monte Carlo values of 5 \mevcsq\ in both cases. In
Fig.~\ref{fig:fgdmcomb} (a), the narrow resonance corresponds to a signal area
of $(25.5\pm 6.5)$ events at a mass difference $\Delta M\,^{+}=(107.8\pm 1.7)
$ \mevcsq\ with a statistical significance of 3.9$\sigma$ . Similarly, a fit to
Fig.~\ref{fig:fgdmcomb} (b) yields a signal area of $(28.0 \pm 7.1)$ events 
at a mass difference $\Delta M\,^{0}=(107.0\pm 1.4)$ \mevcsq\ with statistical 
significance of 3.9$\sigma$. We associate these resonances with
the isospin doublet $\Xi _c^{+\prime }$ and $\Xi _c^{0\prime }$.
To rule out the possibility that the signal is due to random background
under the $\Xi_c$ signal, we reconstruct $\Xi_c\gamma$ combinations using 
fake $\Xi _c$ candidates from the
side-band of the $\Xi _c$ nominal mass region.  The corresponding  mass 
difference distributions $(\Delta M)$ show no evidence of peaking in the 
region of interest. 

In order to probe the systematic stability
of the measured mass differences, we studied the effect of different
background shapes, alternate selection criteria, and the calibration
of the calorimeter absolute energy scale. The major contributor
to systematic shifts was found to be the removal of the $\pi^0$ veto. This has 
the effect of increasing the efficiency by 30\% and 60\% for $\Xi_c^{+\prime}$ and 
$\Xi_c^{0\prime}$, respectively, but also doubling the background, dominantly 
from $\Xi_c^{*}\rightarrow\Xi_c \pi^0$ in which one of the photons from 
$\pi^{0}$ decay is ignored in the reconstruction.
  Based on all these studies we assign
a systematic error to the mass differences of $\pm 2.5$ MeV/$c^2$.

\begin{figure}[ht]
\unitlength 1.0in
\begin{picture}(6.0,6.0)(0.0,0.0)

\put(0.0,0.0){\psfig{file=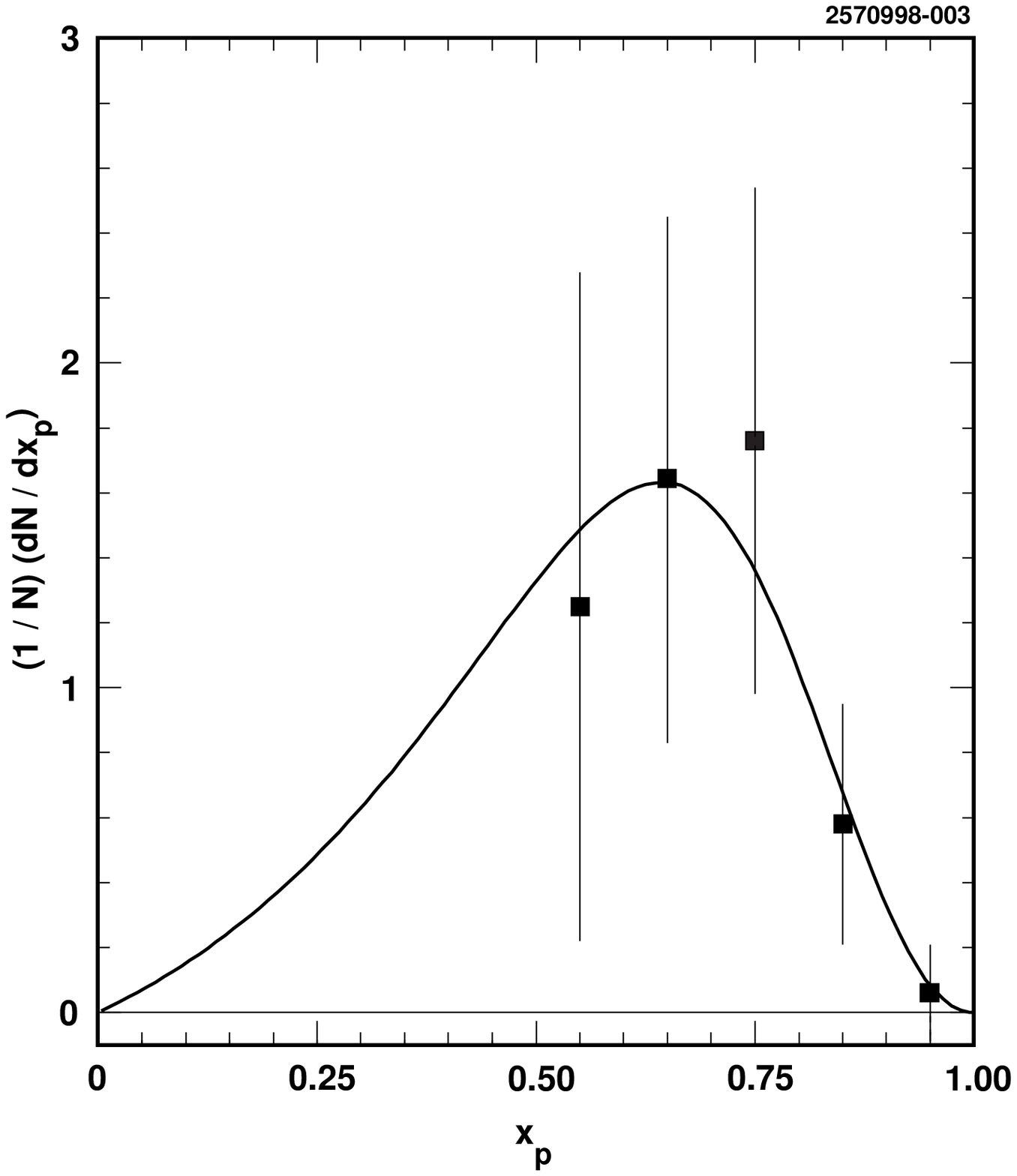,height=5.7in, bbllx=40,bblly=60,bburx=550,bbury=630}}
\end{picture}
\caption[]{ Fragmentation function for $\Xi_c^{\prime}$ (weighted average of 
$\Xi_c^{+\prime}$ and $\Xi_c^{0\prime}$ momentum distributions). }
\label{fig:fgdndxp}
\end{figure}

To measure the $x_p^{\prime}$ spectrum for $\Xi _c^{\prime}$ production, we
assume that at the level of statistics available in our data, the
fragmentation functions for $\Xi _c^{+\prime }$ and $\Xi _c^{0\prime}$
are the same, so that we can combine the data for the two resonances together.
The yield is then obtained as a function of $x_p^{\prime}$ for all the decay modes of
both the resonances from $0.5 < x_p^{\prime} < 1.0$ and corrected for $x_p^{\prime}$-dependent 
reconstruction efficiencies.  The normalized distribution is shown in 
Fig.~\ref{fig:fgdndxp}.  A fit to the Peterson fragmentation 
function\cite{peterson} yields the fragmentation parameter 
$\epsilon _q=0.20_{-0.09}^{+0.23}\pm0.07$, 
which is similar to the previously published result of $\epsilon
_q=0.23_{-0.05}^{+0.06}\pm 0.03$ for $\Xi _c^{+}$ production\cite{edwards}.

We measure that $(37\pm11\pm7)\%$ of all $\Xi_c^{+}$ produced from the 
continuum are from $\Xi_c^{+\prime}$ decays, while $(35\pm9\pm7)\%$ of
all $\Xi_c^{0}$ are from $\Xi_c^{0\prime}$ decays. 
The comparable fraction of  
$\Xi _c^{+}$s from $\Xi _c^{*0}$ decays is $(27\pm 6\pm 6)$\%\cite{avery95}. 
The fraction of $\Xi_c$ from $\Xi_c^{\prime}$ is predicted by Adamov and 
Goldstein\cite{adamov96} to be 1.7 times that from $\Xi_c^*$.  

In conclusion, we have observed two narrow resonances decaying to $\Xi
_c^{+}\gamma $\ and $\Xi _c^0\gamma $. The mass differences $M(\Xi
_c^{+}\gamma \ )-M(\Xi _c^{+})$ and $M(\Xi _c^0\gamma )-M(\Xi _c^0)$ are
measured to be $(107.8\pm 1.7\pm 2.5)$ \ and $(107.0\pm 1.4\pm 2.5)$ \mevcsq, 
respectively; the second error in each case is systematic. This is in
good agreement with theoretical expectations for these mass differences, 
assuming the resonances to be $\Xi_c^{+\prime}$ and $\Xi_c^{0\prime}$, 
respectively.  This is also in good agreement with the models which predict the
mass difference $M(\Xi_c^{*})-M(\Xi_c^{\prime})$ to be about 60-70 \mevcsq.
Since the $J^P=(\frac 32)^{+}$\ charmed strange baryons $\Xi _c^{*+}$\ and $%
\Xi _c^{*0}$\ have already been observed, the most likely interpretation of
the observed resonances\ would be as the $J^P=(\frac 12)^{+}$ charmed
strange baryons\ $\Xi _c^{+\prime }$\ and $\Xi _c^{0\prime }$,
respectively. 

We gratefully acknowledge the effort of the CESR staff in providing us with
excellent luminosity and running conditions.
This work was supported by 
the National Science Foundation,
the U.S. Department of Energy,
Research Corporation,
the Natural Sciences and Engineering Research Council of Canada, 
the A.P. Sloan Foundation, 
the Swiss National Science Foundation, 
and the Alexander von Humboldt Stiftung.



\begin{thebibliography}{99}

\bibitem{alam}  CLEO Collaboration, M. S. Alam \etal, Phys. Lett. B \textbf{226}, 401 (1989).

\bibitem{avery} CLEO Collaboration, P. Avery \etal, Phys. Rev. Lett. \textbf{62}, 863 (1989).

\bibitem{biagi} WA62 Collaboration, S. Biagi \etal, Phys. Lett. B \textbf{150}, 230 (1985)\newline
S. Biagi \etal, Z. Phys. C \textbf{28}, 175, (1985).

\bibitem{coteus}  E687 Collaboration, P. Coteus \etal, Phys. Rev. Lett. \textbf{59}, 1530 (1987).

\bibitem{barlag} ACCMOR Collaboration, S. Barlag \etal, Phys. Lett. B \textbf{236}, 495 (1990).

\bibitem{albrecht} ARGUS Collaboration, H. Albrecht \etal, Phys. Lett. B \textbf{247},
121 (1990).

\bibitem{notation}
For the sake of 
brevity, we use $\Xi_c$, $\Xi _c^{\prime}$, and $\Xi _c^{*}$ to imply 
either the charged or neutral member of the isospin multiplet.  Charge 
conjugates are implied throughout the paper.
\bibitem{avery95} CLEO Collaboration, P. Avery \etal, Phys. Rev. Lett. \textbf{75},
4364 (1995).
\bibitem{gibbons96} CLEO Collaboration, L. Gibbons \etal, Phys. Rev. Lett. \textbf{77},
810 (1996).
\bibitem{franklin}  J. Franklin, Phys Rev. D \textbf{53}, 564 (1996). 

\bibitem{boyd}  C. Glenn Boyd, M. Lu, and M. J. Savage, Phys Rev. D \textbf{55}, 5474 (1997). 
\bibitem{maltman}
K. Maltman and N. Isgur, Phys. Rev. D {\bf 22}, 1701 (1980).
\bibitem{pirjol}  J. G. K\"{o}rner, M. Kr\"{amer} and D. Pirjol, Prog. Part.
Nucl. Physics \textbf{33}, 787 (1994). 
\bibitem{ron}  R. Roncaglia, D. B. Lichtenberg, and E. Predazzi, Phys. Rev. D
\textbf{52}, 1722 (1995). 
\bibitem{savage2}  M. J. Savage, Phys. Lett. B \textbf{359}, 189, (1995). 
\bibitem{falk}  
A. Falk, Phys. Rev. Lett \textbf{77}, 223 (1996).
\bibitem{jenkins}  E. Jenkins, Phys. Rev. D \textbf{54}, 4515 (1996),
 E. Jenkins, Phys. Rev. D \textbf{55}, 10 (1997).
\bibitem{kubota} CLEO Collaboration, Y. Kubota \etal, Nucl. Instrum. Methods Phys. Res. Sec. A \textbf{320}, 66 (1992). 

\bibitem{edwards} CLEO Collaboration, K. Edwards \etal, Phys. Lett. B \textbf{373}, 362
(1996).
\bibitem{peterson}
C. Peterson \etal, Phys Rev. D {\bf 27}, 105 (1993).
\bibitem{adamov96}
A. Adamov and G. Goldstein, hep-ph/9612443, Tufts University, Medford, MA 02155, 20 Dec. 1996.
\end{thebibliography}
\end{document}